\documentstyle [11pt,twocolumn]{report} 
\def\ZzZ{{\hbox{\tenrm Z\kern-.31em{Z}}}} 
\def\CcC{{\hbox{\tenrm C\kern-.45em{\vrule height.67em width0.08em depth- 
.04em 
\hskip.45em }}}}

\newcommand{\bc}{\begin{center}} 
\newcommand{\ec}{\end{center}} 
\newcommand{\be}{\begin{equation}} 
\newcommand{\ee}{\end{equation}} 
\newcommand{\bea}{\begin{eqnarray}} 
\newcommand{\eea}{\end{eqnarray}} 
\newcommand{\bs}{\begin{subequations}} 
\newcommand{\es}{\end{subequations}} 
\newcommand{\beq}{\begin{eqalignno}} 
\newcommand{\eeq}{\end{eqalignno}} 
\def\bol#1{{\bf #1}}

\begin{document} 
%
 
\thispagestyle{empty} 
 
\bc 
\Huge{Quantum dissipation and Neural Net Dynamics}  
 
\vspace{1.2cm} 
 
\large{Eliano Pessa${}^{a}$ and Giuseppe Vitiello${}^{b}$} \\ 
\small 
\bigskip 
{\it ${}^{a}$Facolt\`a di Psicologia, Universit\`a di  
Roma "Sapienza"}\\ 
{\it Via dei Marsi 78, 00185 Roma, Italia}\\ 
{\it and ECONA, Interuniversity Center for Research}\\ 
{\it on Cognitive Processing in Natural and Artificial Systems}\\ 
{\it Roma, Italia}\\ 
{\it ${}^{b}$Dipartimento di Fisica, Universit\`a di Salerno} \\ 
{\it 84100 Salerno, Italia}\\ 
{\it and INFM Unit\`a di Salerno} \\
{\it pessa@axcasp.caspur.it}\\
{\it vitiello@physics.unisa.it }\\ 
\vspace{1.3cm}

\ec 
\small 
{\bf Abstract} 
Inspired by the dissipative quantum model of brain,  
we model the states 
of neural nets in terms of collective modes by the help of the  
formalism of Quantum Field Theory. 
We exhibit an explicit neural net  
model which allows to memorize a sequence of several informations  
without reciprocal destructive interference, namely we solve the  
overprinting problem in such a way last registered information does  
not destroy the ones previously registered. Moreover, the net  
is able to recall not only the last registered information 
in the sequence, but also anyone of those previously registered.

\vspace{1.3cm} 
\normalsize

The quantum model of brain by Umezawa and Ricciardi \cite{UR,S1,S2,CH}  
has attracted much 
attention in recent years.  Moreover, its extension to dissipative 
dynamics \cite{VT}, aimed to solve the long standing problem of memory 
capacity, provides an interesting framework to study consciousness 
related mechanisms. On the other hand, computational neuroscience mostly 
relies on specific activity of neural cells and of their networks, thus 
leading to a number of models and simulations of the brain activity in 
terms of neural nets, mostly based on modern methods of statistical
mechanics and of spin glass theory \cite{MZ,AI}.  
Besides, there is an increasing interest in the 
study of quantum features of network dynamics, either in connection with 
information processing in biological systems, or in relation with a 
computational strategy based on the system quantum evolution (quantum 
computation).  
 
Inspired thus by the papers \cite{UR,S1,S2} and \cite{VT}  
(see also \cite{PR,P2,YA}),  
we explore the possibility of modeling the states  
of neural nets in terms of collective modes by the help of the  
formalism of Quantum Field Theory (QFT). 
 
We show that the classical limit of the 
dissipative quantum brain dynamics (DQBD) \cite{VT}  
provides a representation of a 
neural net characterized by long range correlations among the net's 
units. In this way we exhibit a link between DQBD and neural net 
dynamics \cite{PV}.  
 
We present an explicit neural net  
model which allows to memorize a sequence of several informations  
without reciprocal destructive interference, namely we solve the  
overprinting problem, i.e. last registered information does  
not destroy the ones previously registered. The net  
is also able to recall informations registered prior to the last registered  
one in the sequence. 
 
In the following we will first introduce the general  
theoretical background on  
which our neural net is modeled and then we will present some 
of its specific features and 
the results, which, although preliminary, confirm our expectations. 
 
We consider a three-dimensional set of $N$ interacting units (neural  
units) sitting each one in a space-time site 
$x_{n} \equiv ({\bol x_{n}}, t_{n}), n = 1,2,..N$. Each unit can be in  
the state $on~ (1)$ or $off~ (0)$. 
The neural unit activity is characterized by the amplitude of the emitted  
pulse and by the phase determined by the emission time. This suggests to us  
that each unit can be described by a complex doublet field  
$\psi(x_{n}) = (\psi_{u} (x_n),~ \psi_{d} (x_n))$, 
with $\psi_{u}(x_{n})$ and $\psi_{d}(x_{n})$ complex field  
components, $u$  
and $d$ denoting the field inner degrees of freedom corresponding to $on$  
and $off$, respectively. 
 
At each site $x_n$ the field inner variable may assume a well specified  
value ($u$ or $d$). The set of these values for all the sites specifies the  
{\it microscopic} configuration in the 
$(u-d)$-space; however, in full generality the specification of the  
{\it macroscopic or functional state} of the net does not actually requires  
that correspondingly one should have a unique, well definite microscopic  
configuration where each unit state is specified by a definite $u$ or $d$  
value. In general, indeed, many distinct microscopic configurations of the  
component units may correspond to the same functional state of the net.  
 
This means that a given (equilibrium) state of the net may be well  
compatible with fluctuations in the states of the individual component  
units. This amounts to say that the net state is not strictly and crucially  
dependent on the specific state of each individual unit: i.e. we admit  
enough {\it plasticity} (as contrasted with {\it rigidity}) for  
the net; in other words, we can say that the net macroscopic state is  
the output, or the {\it asymptotic} state, emerging from the microscopic  
dynamics which rules the interaction among the component units.  
 
For large number $N$ of component units, such a picture is certainly more  
"realistic" than a "rigid" one and could also be more appropriate for a  
possible modeling of the natural brain in terms of neural nets (as it is  
well known the brain functional activity is not strictly related with the  
activity of each single neuron; in this paper however we will not deal  
with modeling the natural brain). 
 
One can view such a situation also from the  
perspective of the pulses or signals traveling on the connections among  
the units: a traveling signal may contribute to excite or de-excite a  
certain unit thus changing its $u$ or $d$ state. Consequently, a  
specific state of the net, corresponding to a  
given {\it dynamical} distribution of pulses on the net connections, is  
necessarily a state for which the single unit states at each site cannot  
be uniquely specified once for ever, due to pulse action on the units. As  
a matter of fact, one should consider the unit states as non-observable,  
since any observation on the unit may non-trivially interfere with the  
dynamics of the pulses. Only the output of such a dynamics is observable  
and this is why above we have called it the asymptotic state of the 
net (i.e. states for time $t_n \rightarrow \pm \infty$ for each $n$). 
 
Summing up, since fluctuations are allowed for  
the states of the individual unit at each site, and thus for the basic  
field $\psi (x_n)$, and since, as a consequence, in the ($u-d$)-space the  
uncertainty in the identification of the "trajectory" representing the  
evolution of the state of the unit cannot be eliminated without strongly  
interfering with it, we are led to treat $\psi (x_n)$ like
a quantum field satisfying quantum dynamical equations.  

The above considerations lead us to think of the neural net in terms  
similar to the ones usually adopted for condensed matter physics: the  
global behavior of the net, namely its functional state and its  
evolution, can be characterized by a (classical) macroscopic observable  
as it usually happens in solid state physics, e.g. in superconductivity,  
in ferromagnetism, etc.. Such an observable, generally called the ``order  
parameter'', is determined by the dynamics of the elementary components  
or units and by its symmetries \cite{TFD,U2,IZ,A1}. 
It may be considered as  
a {\it code} specifying the vacuum or ground state \cite{UR,VT}. 
 
Like in ferromagnetism one introduces the order  
parameter "magnetization", in our present case we introduce the  
macroscopic observable $\cal M$ whose values are assumed to specify the  
information content of the net. We define ${\cal M}  
\equiv (1/2){|(N_u - N_d)|}$, with $N = N_u + N_d$, 
where $N_u$ and $N_d$ denote the number of units $on$ and $off$,  
respectively. $\cal M$ is the neural net order parameter which  
characterizes its macroscopic state. 
 
The state ${\cal M} =0$ is called the "normal state" (void of information  
content); the "information states" or ``memory states''are the ones with
${\cal M} \neq0$   
(different informations associated to different non-zero $\cal M$  
values).  
 
Since the information comes to the net {\it from the outside}, we assume  
that the neural net may be set into a ${\cal M} \neq 0$ state under the  
action of an external input (coupling of the net with the environment)
and it remains in such a state even after the external input is not
anymore acting on the net (the information has been recorded). In other
words, we assume that the interaction among the net units cannot force, by  
itself, the net into a ${\cal M} \neq 0$ state (i.e. prior of any
external input the net remains in its normal state). This in turn means  
that, from one side, the basic dynamics describing the interaction among  
the units (i.e. the evolution equations for the basic $\psi (x_n)$ field)  
must be invariant under the $SU(2)$ group of transformations acting on  
the doublet field $\psi (x_n)$. On the other side, it also means that the  
ground state is {\it not} invariant under the full $SU(2)$ group of  
transformations, i.e. the external input triggers the spontaneous
breakdown of the $SU(2)$ symmetry.
 
Notice that, as observed above, there can be many configurations of the  
set of units (microscopic configurations) corresponding to a given value of  
$\cal M$, and therefore to a given state of the net: $\cal M$ specifies  
indeed only the difference $|(N_u - N_d)|$, but says nothing on  
which ones 
and how many are the sites $u$ and which ones and how many are those $d$;  
so that any change, or fluctuation, between the $u$ and $d$ state of the  
units in different sites is allowed, provided the difference $\cal M$ is  
kept constant. In this sense $\cal M$ is a {\it macroscopic}  
variable. On the contrary, the $\psi (x_n)$ fields determine the {\it  
microscopic} configurations.  
 
Moreover, the quantum field dynamics  
generates asymptotic equilibrium states with negligible fluctuations  
of $\cal M$. The macroscopic  
"memory" state of the neural net indexed by $\cal M$ is then a {\it  
classical limit state} in the sense of QFT, namely the state for which the  
fluctuations in the number of certain modes is negligible with respect to  
the number of the same kind of modes {\it condensed} in it: in other words,  
a {\it coherent} state \cite{KS} with respect to these modes. 
These condensed modes are long range correlation modes. 
 
We have in conclusion two levels of description: i) The dynamical level 
and ii) the asymptotic level. At the dynamical level the interaction  
among the neural units (represented by basic fields $\psi (x_n)$)  
is ruled by a  
certain set of dynamical equations, which are assumed to be invariant 
under the $SU(2)$ group;  this level is precluded to  
observations. At the asymptotic level, the $SU(2)$ symmetry is  
spontaneously broken and the neural net state is  
characterized by the order parameter $\cal M$. 
 
The set of asymptotic fields includes the field describing 
the non-interacting, 
"free" (at $t_n \rightarrow \pm \infty$ for each $n$) units, 
say $\phi (x_n)$, which also is a complex doublet field with 
($u-d$) inner freedom, and other fields which are generated by the  
dynamics. By resorting to well known results in QFT \cite{IZ,TFD}, 
indeed, whenever the order parameter $\cal M$ is different from  
zero, the dynamics generates excitation fields describing {\it long range  
correlations} among the units, which are therefore {\it collective modes} 
(the Goldstone theorem). These  
long range modes are massless and thus their condensation in the ground
state does not change its energy; it only produces other (ground) states
degenerate in the energy, different among themselves for their
condensation content. The stability of the memory states is thus insured.
The value of the order parameter $\cal  
M$ is a measure of the collective mode condensation in the ground state. 
 
Since the net is an open system (coupled with the environment) and the
information storage produces by   
itself the breakdown of the time-reversal symmetry \cite{VT}, we consider  
the QFT for dissipative systems and the neural net state can be then 
recognized \cite{PV} to be a finite temperature state of QFT. 
 
The role of dissipation is crucial in solving the {\it  
overprinting} problem, namely the problem of the net memory capacity:  
in a sequential information recording each information storage would  
over-impose itself to the previously recorded one, thus deleting it. 
In the dissipative dynamics, on the  
contrary, a large memory capacity is possible since 
each information is recorded in each of the many degenerate ground states,  
without destructive interference among them
\cite{VT}. 
 
In other words, dissipation implies that the net overall
state may be represented as a superposition 
of infinitely many degenerate ground states, or memory states, each of them 
labeled by a different code number and each of them independently  
accessible to information storage. Many information "files" may then 
coexist thus allowing a huge memory capacity. 
Non-unitary equivalence among different memory states
acts as a protection against overlap or  
interference among different informations \cite{VT}. 
 
In realistic neural net, the finiteness of the "volume" (the number of  
neural units) and possible defect effects may spoil unitary  
non-equivalence thus leading to information interferences and distortions. 
 
The retrivial of information is described by "reading off" the mirror  
modes of the same code number of the information to be recalled. These  
mirror modes are essentially a "replication signal" of the one  
responsible for memory storage. The replication signal thus acts as  
a probe by which one "reads" the stored information. The process of
information recalling drives the net into the (macroscopic) memory state
of code ${\cal M}$ corresponding to that specific information to be
retrived. 
 
We also observe that the mirror modes may acquire an effective nonzero  
mass due to the effects of the system finite size. 
Such an effective mass then 
introduces a threshold in the energy to supply in order to trigger  
the "recall" process. This may lead, from one side, to "difficulties"  
in the information retrivial; on the other side, it may act as a  
"protection" against unwanted perturbations and cooperate to the neural  
net memory state stability.  
 
The study of thermodynamic properties shows that the generator  
of time evolution of the net state is the system entropy. The stationarity  
of free energy implies the Bose distribution for the collective modes and  
the Fermi distribution for the neural unit fields. In this way the  
traditional activation function for neural net units is recovered. 
For further details see \cite{VT} and \cite{PV}. 
 
Let us now briefly present some of the features of the  
specific neural net model we have worked out and the related results. 
We will give here only a qualitative description of the model. See  
\cite{PV} for formal details. 
 
The net dynamics is given by the Pauli-Dirac equation for a doublet  
field $\psi$ interacting with an external magnetic field representing  
the external input. A spatial discretization of this equation on a two
dimensional lattice of 
$20 \times 20$ sites is adopted, so as to transform the original field
equation 
into a system of coupled ordinary differential equations. 
Besides the interaction with the external magnetic field, the  
doublet field also interacts with the mean magnetic field  
which is generated over  
the net as a response to the external input.  
We use the Bragg-Williams approximation and the 
mean magnetic field is taken to be proportional to  
the magnetization induced by the external  
input with a proportionality constant given by the Weiss constant 
$\gamma$. The dynamical equation thus presents a nonlinear term in  
$\psi$ since the magnetization is by itself given in terms of a  
bilinear form in $\psi$. The Weiss constant $\gamma$ is also related  
to the mean value over the whole net of the mean values of the 
connection strength for each site. 
 
The practical implementation of the net was done through  
the following steps:

a) after explicit representation of the $\psi$ field components 
$\psi_{u}$ and $\psi_{d}$ in terms of their real and imaginary  
parts, we obtained the corresponding four equations  
from the Pauli-Dirac equation. 

b) each of the four component field variables was expressed as a   
product of the site activation function times the connectivity 
potential of the site itself. 

c) we considered the nearest neighbor approximation for the site  
connections. 

d) after a spatio-temporal discretization, we associated to each
unit (i.e. to each site) a sigmoidal activation function characterized
by a ``temperature'' parameter.  

e) we assumed an independent evolution for each of the four component 
field variables. 

f) we used the  modulus squared value of the activation function of
each unit to determine the time evolution of the activation dynamics of the
unit itself.

g) we used simulated annealing in each process of writing and of  
reading (recalling). 
 
Through the implementation of the above steps we obtained a neural net
able to record
a sequence of informations without  
overprinting (i.e. without destruction of previously registered  
informations in the course of a subsequent registration process)  
and to able to recall anyone of the registered informations (i.e.  
not simply the last one) under presentation of an external input  
similar to the one to be recalled. 
 
Such results make us confident that a novel conceptual and  
formal scheme in neural net modeling may be introduced which is  
based on the simulation of a quantum dynamical evolution. 

We are glad to aknowledge partial support from MURST and INFM. 
 

\end{document}